\documentclass[reprint,prc,groupedaddress,showpacs,preprintnumbers,amsmath,amssymb]{revtex4-1}
\usepackage{graphicx}
\usepackage{dcolumn}
\usepackage{amsmath}
\usepackage{times}

\begin{document}

\title{ The $\alpha-\alpha$ fishbone potential revisited }

\author{J.\ P.\ Day}\altaffiliation{Present address: Institute for Theoretical Physics, University of Graz,  A-8010 Graz, Austria}
\author{J.\ E.\ McEwen}\altaffiliation{Present address: Department of Physics, The Ohio State University, Columbus OH, 43210}
\author{M.\ Elhanafy}
\author{E.\ Smith}
\author{R.\ Woodhouse}
\author{Z.\ Papp}
\affiliation{ Department of Physics and Astronomy,
California State University Long Beach, Long Beach, California, USA }

\date{\today}

\begin{abstract}
The fishbone potential of composite particles 
simulates the Pauli effect by nonlocal terms. We determine the $\alpha-\alpha$ fishbone potential
by simultaneously fitting to two-$\alpha$ resonance energies, experimental phase shifts and three-$\alpha$ binding energies.
We found that essentially a simple gaussian can provide a good description of two-$\alpha$ and three-$\alpha$ experimental data
without invoking three-body potentials.
\end{abstract}


\maketitle

\section{Introduction}

The potential is the most important non-observable physical quantity in quantum mechanics. 
It is not an observable, yet it determines all the physical observables. There are two ways of determining the quantum 
mechanical potential. If the quantum system has a classical counterpart, one can use the correspondence principle. Otherwise we 
determine the potential by calculating observables and comparing them to experiments. 

Almost all, so called ''elementary'' particles are in fact composite particles made of even more elementary particles. 
These constituents are fermions that obey the Pauli principle, i.e.\ they cannot occupy the 
same quantum state.  The Pauli principle, and the internal structure and dynamics of the constituent 
fermions can lead to a very complicated potential for the composite particles. 

The simplest way to model the Pauli effect is to use a repulsive short range potential, which suppresses the 
wave function at short distances. The parameters of this phenomenological potential are
determined either by an the inverse scattering method or by fitting models to experiments. In most models local potentials are used. 
Generally these models cannot provide an acceptable description of three body data, which results in the need for 
three-body potentials.

An alternative approach is that we try to incorporate all the information about the internal structure and dynamics of
composite particles into their mutual interactions. Several potentials based on the nuclear cluster model \cite{wildermuth} 
have been proposed. One possibility is to derive the interactions from the cluster model in the framework of the resonating group method 
(see eq.\ Ref.\ \cite{theeten} for a recent review). The  other approach is more phenomenological. It uses some information on the structure
of composite particles,  but incorporates some phenomenological potential, whose parameters are determined by fitting the results to experiments.
The pioneer of this type of model is the orthogonality condition model by Saito
\cite{saito}, where the states are orthogonal to predefined Pauli forbidden states. 
In the method of Buck, Friedrich and Wheatley \cite{bfw} a
deep potential is adopted, and it is assumed that the lowest few states are forbidden by the Pauli principle. 
The fishbone model by Schmid \cite{schmid1,schmid2} goes beyond previous models as it introduces the concept of partially Pauli 
forbidden states. 

It was a common belief that if we incorporate 
all the information about the internal structure of the particles into their mutual interactions then the 
three-body potential would be small, perhaps negligible. 
However, for $\alpha$
particles, while phenomenological shallow local potentials under-bind the three-$\alpha$ system, 
cluster model inspired phenomenological potentials considerably over-bind them. 

In this work we revisit the problem of interaction of composite particles.  We consider the fishbone model of the  $\alpha-\alpha$ 
interaction. We have chosen the $\alpha-\alpha$ potential, because the $\alpha$ particle has an exceptionally strong binding energy. 
We adopted the fishbone model because, in our opinion, this is the most elaborated phenomenological
cluster-model-motivated potential. The variant of the fishbone potential has been designed to minimize and  to 
neglect the three-body potential. Therefore we can try to determine the interaction by a simultaneous fit to two-body and three-body data.

Previously we studied the $\alpha-\alpha$ fishbone model \cite{pm} and proposed a new parametrization of the 
fishbone $\alpha-\alpha$ interaction. We fitted the two-body phase shifts and the three-$\alpha$ ground state energy. 
Later we found that the results are not stable against varying the parameters. Here, besides two body phase shifts and the three-$\alpha$ ground state, 
we include the three-$\alpha$ $L=0$ ground, the $L=0$ resonant state, and the $L=2$ bound state.

In Section II we will outline the fishbone model for the composite particles. In Section III we determine the $\alpha-\alpha$ 
potential by using two-$\alpha$ and three-$\alpha$ data. Finally we draw some conclusions.

\section{The fishbone optical model}

The fishbone model is motivated by the cluster model. In the resonating group model the total wave function is an antisymmetrized 
product of the cluster $\Phi$ and the inter-cluster $\chi$ relative states
\begin{equation}
|\Psi \rangle = |\{{\cal A} \Phi \chi \}\rangle.
\end{equation}
The state $\Phi$, which is supposed to be known in this model, describes the internal properties of the clusters, including spin and isospin structure. 
The unknown relative motion state $\chi$ is determined from the variational ansatz
\begin{equation}
\langle \Phi \delta \chi | {\cal A} (H-E) {\cal A} | \Phi \chi \rangle =0.
\end{equation}
This ansatz results in a rather complicated equation for $\chi$, which was possible to solve only by using serious approximations on 
$\Phi$ and on the interaction of the particles. In a typical example $\Psi$ describes fermions in harmonic oscillator 
potential wells located at some distance apart and $\chi$ is the relative motion of the oscillator wells. We can easily see that
if we express $\chi$ in terms of harmonic oscillator states, some of the lowest states in the relative motion space are not allowed 
due to the Pauli principle.

The Schr\"odinger equation for the two-body fishbone model is given by \cite{schmid1}
\begin{equation}
(h^{0}+{\cal V}_{l}) |\chi \rangle = E | \chi \rangle,
\end{equation}
where $h^{0}$ is the kinetic energy operator.
Our knowledge on the internal structure and the Pauli principle are incorporated in the fishbone potential
\begin{equation}
{\cal V}_{l}=v_{l}-\sum_{i,j}| u_{l,i}\rangle \langle u_{l,i}|(h^{0}+v_{l}-\epsilon_{l,i})|u_{l,j}\rangle \bar{M}_{l,ij}\langle u_{l,j}|,
\end{equation}
where $l$ refers to partial wave and $v_{l}$ is a local potential. The states $|u_{l,i}\rangle$ are eigenstates of the norm operator,
\begin{equation}
\langle \Phi \vec{r} | {\cal A} | \Phi u_{l,i} \rangle = (1-\eta_{l,i})\langle \vec{r} |u_{l,i}\rangle,
\end{equation}
where $\vec{r}$ is the center of mass distance of the two clusters. If the relative motion is forbidden by Pauli principle then $\langle \Phi \vec{r} | {\cal A} | \Phi u_{l,i} \rangle=0$,
and $\eta_{l,i}=1$. The $\eta_{l,i}$ eigenvalues are ordered such that $|\eta_{l,i}| \ge |\eta_{l,i+1}|$. The matrix $\bar{M}$ is then given by
\begin{equation}
\bar{M}_{ij}=
\begin{cases}
1-\frac{ \displaystyle 1-\eta_{l,i}}{ \displaystyle [(1-\bar{\eta}_{l,i}) (1-\bar{\eta}_{l,i})]^{1/2}}, & \text{if $i \le j$}, \\
 1-\frac{ \displaystyle  1-\eta_{l,j}}{ \displaystyle   [(1-\bar{\eta}_{l,j}) (1-\bar{\eta}_{l,i})]^{1/2}}, & 
 \text{if $i > j$},
 \end{cases}
\end{equation}
where $\bar{\eta}_{l,i}=0$ if $\eta_{l,i}=1$ and $\bar{\eta}_{l,i}=\eta_{l,i}$ otherwise.  In matrix form, if we have one Pauli forbidden state, we have
\begin{equation}
\bar{M}_{l}=\left(\begin{matrix}
1 & 1 & 1 & 1 & \ldots \\
1 & 0 & 1-\sqrt{\frac{1-\eta_{l,2}}{1-\eta_{l,3}}} & 1-\sqrt{\frac{1-\eta_{l,2}}{1-\eta_{l,4}}} &  \ldots \\
1 & 1-\sqrt{\frac{1-\eta_{l,2}}{1-\eta_{l,3}}} & 0 & 1-\sqrt{\frac{1-\eta_{l,2}}{1-\eta_{l,4}}} &   \ldots \\ 
1 & 1-\sqrt{\frac{1-\eta_{l,2}}{1-\eta_{l,3}}} & 1-\sqrt{\frac{1-\eta_{l,2}}{1-\eta_{l,4}}} & 0 &   \ldots \\ 
\vdots & \vdots & \vdots &\vdots & \ddots  \\
\end{matrix}\right).
\end{equation}
The matrix elements of $\bar{M}$ exhibit a fish-bone-like structure; hence the name of the model. 
In this model the Pauli-forbidden states become eigenstates at $\epsilon$ energy. 
By choosing $\epsilon$ as large positive, they become bound states at large positive energy, and thus disappear from 
the physically relevant part of the spectrum. 
There are several versions of the fishbone model which differ in off-shell transformations, i.e.\ in transformations 
which effect the internal part of the wave function and leave the asymptotic part, and consequently the spectrum, unchanged. 
This version of the model minimizes the three-body potential.

We can extend the two-body fishbone model to three clusters by embedding the two-body fishbone 
potential into the three-body Hilbert space \cite{schmid2}. We use the usual configuration space Jacobi coordinates. For example,
the coordinate $x_{1}$ denotes the vector between particles $2$ and $3$, while $y_{1}$ connects the center of mass of the 
subsystem $(2,3)$ with the particle $1$.
The three-body fishbone Hamiltonian is given by
\begin{equation}
H = H^{0} + {\cal V}_{x_{1}} {\bf 1}_{y_{1}} + {\cal V}_{x_{2}} {\bf 1}_{y_{2}} + {\cal V}_{x_{3}} {\bf 1}_{y_{3}},  
\end{equation}
where $H^{0}=h^{0}_{x_{i}}+h^{0}_{y_{i}}$, with $i=1,2,3$, is the kinetic energy operator. Here we have omitted the three-body
potential.

The fishbone potential is rather complicated. It has a local Coulomb-like part augmented by a nonlocal short range potential. 
The numerical treatment is also nontrivial. 
However, in the past couple of years, in a series of publications, we developed a method for 
dealing with potentials of this type. We solve the Lippmann-Schwinger integral equation for two-body problems and Faddeev integral 
equations for three-body problems. We approximate the short-range parts of the potential in the Coulomb-Sturmian basis. This basis
allows an exact analytic evaluation of the Coulomb Green's operator,
in terms of a continued fraction for the two-body case, and in terms of a contour integral
for the three-body case. For details, see eg.\  Ref.\ \cite{pm} and references therein.

\section{The fishbone model of $\alpha-\alpha$ interaction}

We adopt a model that in the $\alpha$ particles the nucleons are in $0s$ states in a harmonic oscillator well of width parameter $a$. 
The norm 
kernel eigenvalues are also harmonic oscillator states with the same width parameter and the eigenvalues are known
\cite{horiouchi}: $\eta_{0,i}=1,1,1/4,1/16,1/64,\ldots$, $\eta_{2,i}=1,1/4,1/16,1/64,\ldots$ and $\eta_{4,i}=1/4,1/16,1/64,\ldots$.  
This shows that in the 
$l=0$ relative motion channel there are two Pauli-forbidden states, in $l=2$ there is one, and 
in $l=4$ and higher channels there are none. 
The decreasing value of $\eta$ indicates that in the relative motion the harmonic oscillator sates with higher quantum number  
are less and less suppressed by the Pauli principle. 
For the $\epsilon$ parameter of the fish-bone model, which aims  to remove the Pauli-forbidden states, we took 
$\epsilon=60000$ MeV. In this range of $\epsilon$, the dependence of the results was beyond the fifth significant digit. 
We used the experimental phase shifts from Ref.\ \cite{alibodmer}.

A fishbone potential of the $\alpha-\alpha$ system was determined by  Kircher and Schmid \cite{kircher}. 
The harmonic oscillator width parameter was fixed to $a=0.55 \; \mbox{fm}^{-2}$, which leads to the length parameter 
$r_{0}=(2a_{0})^{-1/2}=0.9535\: \mbox{fm}$. The local potential was taken in the form
\begin{equation}\label{locpot}
v_{l}(r)=v_{0} \exp(-\beta r^{2})+\frac{4\mbox{e}^{2}}{r} \mbox{erf}\left( \sqrt{\frac{2a}{3}} r \right),
\end{equation}
where $v_{0}$ and $\beta$ are fitting parameters. They were determined by fitting to experimental phase shifts.
The values  $v_{0}=-108.41998\; \mbox{MeV}$ and $\beta= 0.18898\; \mbox{fm}^{-2}$ were found to provide the best fit.

While this potential provides a reasonably good fit to $l=0$ and  $l=2$  phase shifts, it results in
too much binding for the three-$\alpha$ system. It gives $E^{3\alpha}_{L=0}\sim -15.5$ MeV, 
while the experimental three-$\alpha$ binding energy 
is $E^{3\alpha}_{exp}=-7.275$ MeV. One may conclude that similarly to conventional local potential models, 
there is a need for a three-body potential. 
This was the choice that Oryu and Kamada \cite{oryu} adopted. 
They added a phenomenological three-body potential to the fish-bone potential of Kircher and Schmid and found that a 
huge three-body potential is needed to reproduce the experimental data. So, the conclusion was, that although the fishbone potential
provides a good fit to two body data, it needs strong three-body force to reproduce the three-body data. 
This is with a the potential which was designed such that the three-body force could be 
neglected. This is certainly not the case. So, although, the model has some good features with this parametrization, 
it do not meet up to its promise.

\section{Results}

\begin{figure}[th]
\centering
\includegraphics[width=8.5cm]{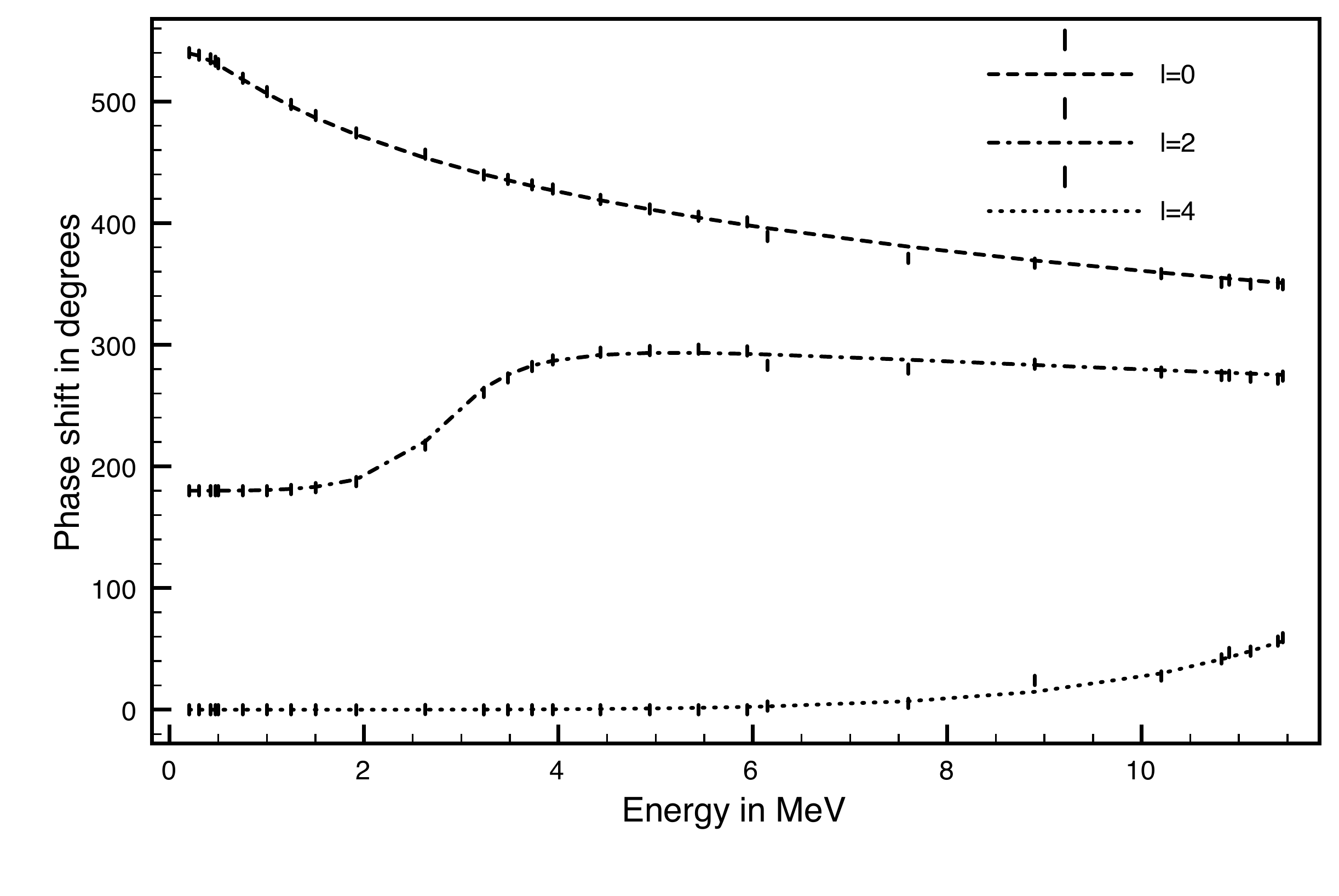}
\vspace*{8pt}
\caption{Fit to the experimental $l=0$, $l=2$ and $l=4$ phase shifts from two-body data.}
\label{fig1}
\end{figure}

We refitted the two-$\alpha$ experimental data with the potential (\ref{locpot}). Besides the phase shift data we also incorporated 
the famous 
$l=0$ two-$\alpha$ resonance state at $E_{\alpha-\alpha}= 0.0916- 0.000003 \mathrm{i}$ MeV. We found that the parameters
$a=0.5838\; \mathrm{fm}^{-2}$, $v_{0}= -109.97\;\mathrm{MeV} $ and $\beta= 0.19417 \; \mathrm{fm}^{-2}$ provide the best fit. 
The fit to the $\alpha-\alpha$ phase shift is given in Fig.\ \ref{fig1}. 
By calculating three-$\alpha$  states we get $E^{3\alpha}_{L=0} = -13.6$ MeV, $E^{3\alpha}_{L=0} = -0.2 $ MeV, and
$E^{3\alpha}_{L=2} = -11.3$ MeV. So, still the fishbone model with parameters extracted from two-body data only, 
cannot provide a good description to three-body experimental binding energies.

On the other hand, two-body data are sensitive mostly on the outer part of the potential and they are less sensitive on the inner part.
We may use this fact to find a fishbone potential which fits simultaneously to two-body and three-body data. 
So, besides the two-body phase shifts and the $l=0$ two-$\alpha$ resonant state, we incorporated the 
$L=0$ three-$\alpha$ ground state at $E= -7.275$ MeV, the $L=0$ three-$\alpha$ excited state at $E=0.375$ MeV, and the $L=2$ 
three-$\alpha$ bound state at $E= -2.836$ MeV.

\begin{figure}[th]
\centering
\includegraphics[width=8.5cm]{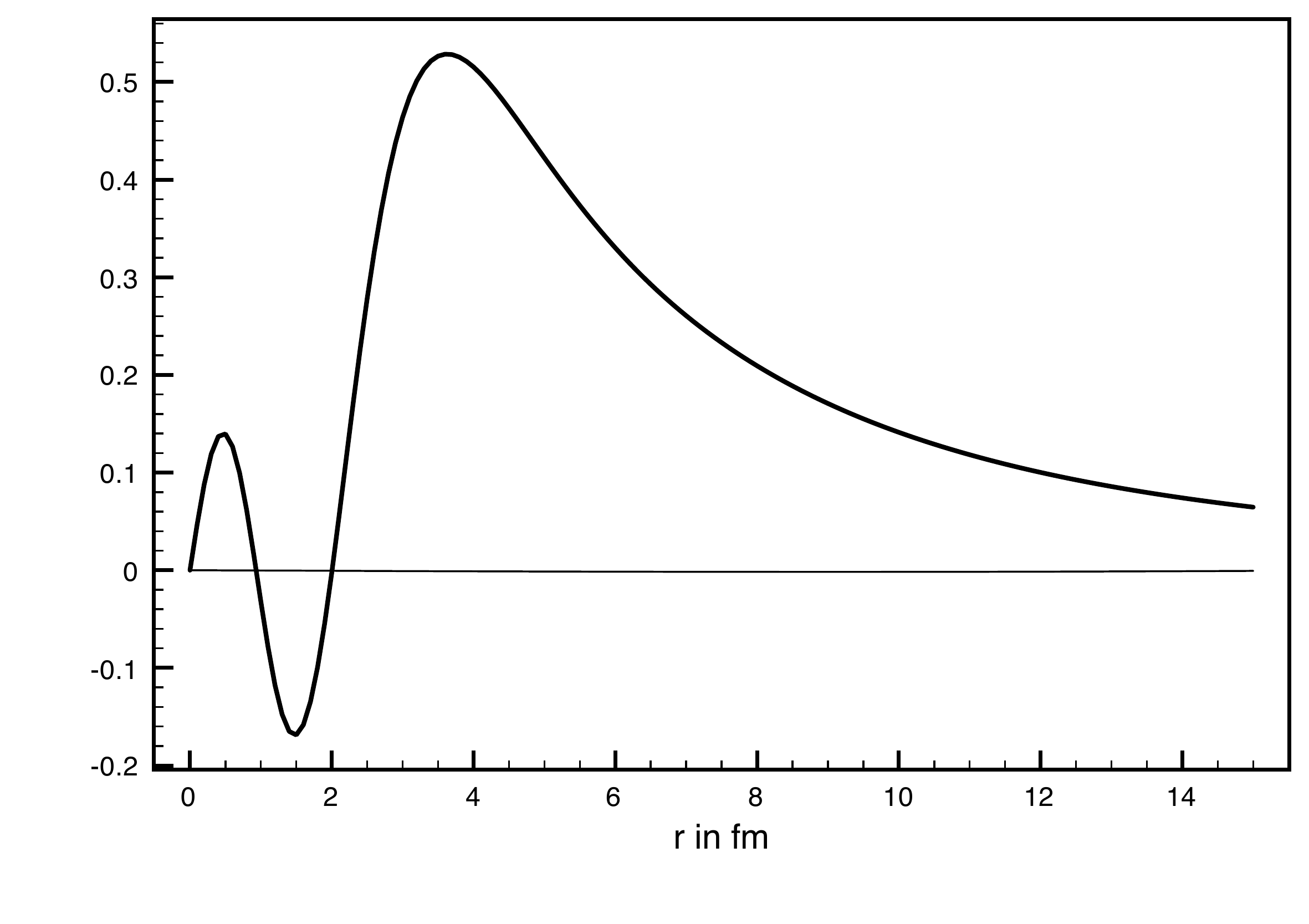}
\vspace*{8pt}
\caption{The wave function of the $l=0$ $\alpha-\alpha$ resonant state at energy $E_{\alpha-\alpha}= 0.09158- 0.000003 \mathrm{i}$ MeV.}
\label{2awf}
\end{figure}

\begin{figure}[th]
\centering
\includegraphics[width=8.5cm]{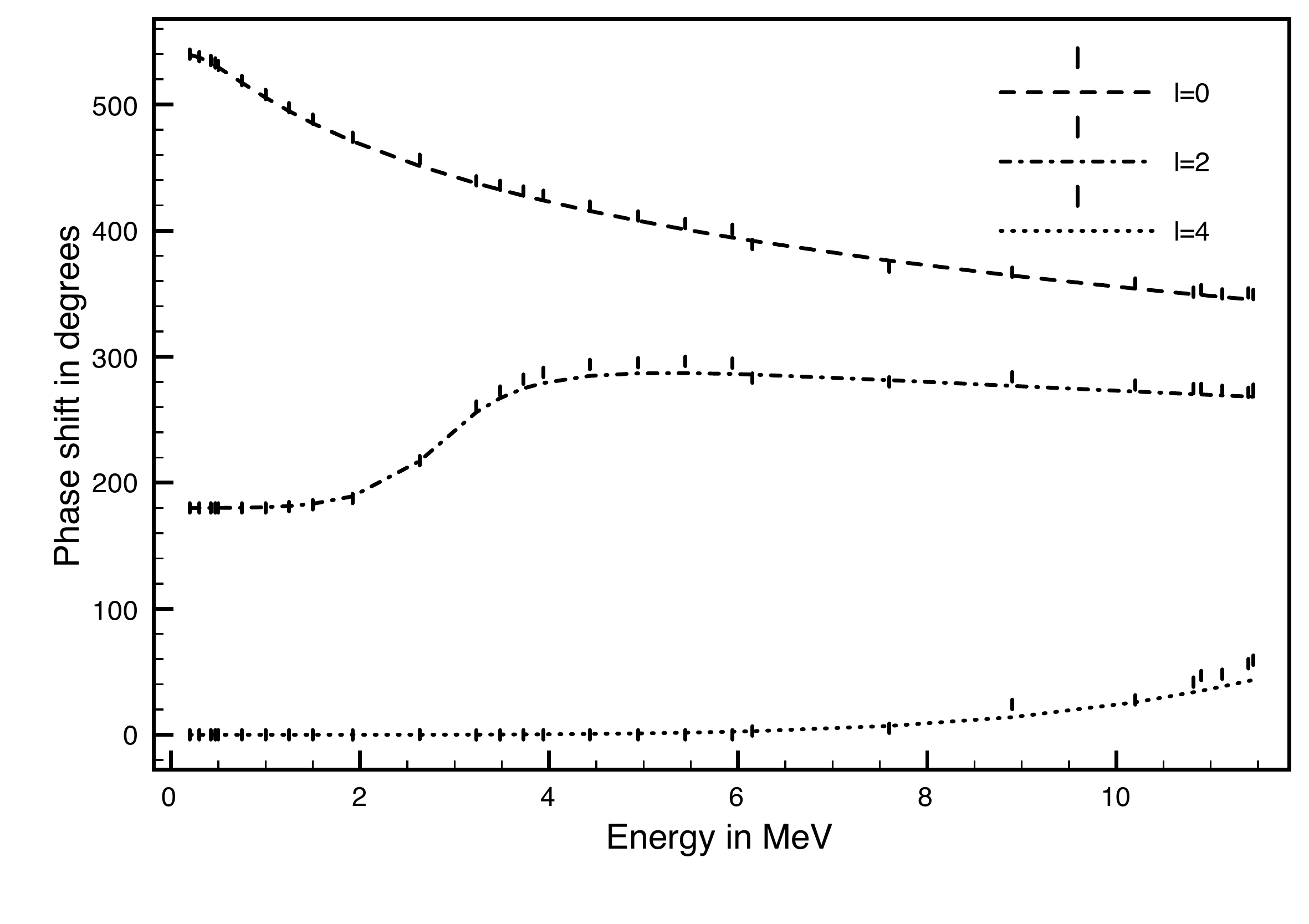}
\vspace*{8pt}
\caption{Fit to the experimental $l=0$, $l=2$ and $l=4$ phase shifts from two-body and three-body data.}
\label{fig2}
\end{figure}

We achieved the best fit to experiments with parameters $a=0.6266\; \mathrm{fm}^{-2}$, $v_{0}= -101.78\;\mathrm{MeV} $
and $\beta= 0.1881 \; \mathrm{fm}^{-2}$. This set of parameters provides an $l=0$ two-$\alpha$ resonant state 
at $E_{\alpha-\alpha}= 0.09158- 0.000003 \mathrm{i}$ MeV. The corresponding wave function is shown in Fig.\ \ref{2awf}.  
Note the peculiarity of the fishbone model in that the ground state 
wave function has nodes due to the orthogonality to the fully Pauli forbidden states.
Fig.\ \ref{fig2} shows the $\alpha-\alpha$ phase shifts. We can see that the agreement with the experiments is almost as good as before.
For the three-$\alpha$  states we get $E^{3\alpha}_{L=0} = -7.01$ MeV, $E^{3\alpha}_{L=0} = 0.51 $ MeV, and
$E^{3\alpha}_{L=2} = -4.5$ MeV. The $L=0$ states are almost in perfect agreement with the experimental values.
The $L=2$ bound state is slightly over-bounded.

\section{Summary and conclusion}

In this work we propose a new parametrization of the fishbone $\alpha-\alpha$ potential.  
We determined the potential by a simultaneous fit to two-$\alpha$ and three-$\alpha$ data. We found that with three
fitting parameters one can describe the two-body resonance and phase shifts in all partial waves. 
The potential also provides a reasonably good 
description to three-body data without invoking any three-body potential.

We can learn from this study that if we incorporate our knowledge on the structure of composite particles  into their interaction, like
we do in the fishbone model,
we can achieve a substantial simplification of the potential. Here, in the $\alpha-\alpha$ case, we have only three parameters, 
while in the conventional Ali-Bodmer-type potentials, we have a couple of  independent parameters in each partial wave, 
plus additional parameters for the three-body potential.
We believe that cluster-model-based models for interaction of composite particles deserve further considerations. 
The fishbone model of Schmid is 
especially appealing as it faithfully represents the Pauli principle and uses the concept of partly Pauli forbidden states. 
It could also serve as a framework for the nucleon-nucleon potential.

\section*{Acknowledgments}

The authors are indebted to E.\ W.\ Schmid  for useful discussions. We gratefully acknowledge support from the 
Army High Performance Computing Research Center consortium "Hispanic Research and Infrastructure Development Program".


\end{document}